\documentclass[usenatbib,psfig,epsfig,preprint]{mn2e}

\usepackage{epsfig}

\title[Discovery of Irradiation Induced Variations in V2275 Cyg]
{Discovery of Irradiation Induced Variations 
in the Light Curve of the Classical Nova Cygni 2001 No.2 (V2275 Cyg)}
\author[\c{S}. Balman et al.]{\c{S}. Balman$^{1}$\thanks{E-mail:
solen@astroa.physics.metu.edu.tr}, A. Y{\i}lmaz$^1$,
A. Retter$^{2}$, T. Sayga\c{c}$^3$, H. Esenoglu$^{3,4}$\\
$^{1}$Department of Physics, Middle East Technical University,
Ankara, 06531, Turkey\\
$^{2}$Department of Astronomy and Astrophysics, Pennsylvania State University,
525 Davey Lab., University Park, PA 16802, USA\\
$^{3}$Istanbul University, Dept. of Astronomy and
Space Sciences, Istanbul, 34119, Turkey\\
$^{4}$Istanbul Observatory, Research and Application Center, Istanbul, 34119, Turkey}

\begin{document}

\date{Accepted . Received ; in original form }


\maketitle

\label{firstpage}

\begin{abstract}

We present the CCD photometry, light curve and time series analysis
of the classical nova V2275 Cyg (N Cyg 2001 No.2). The source was observed
for 14 nights in total using an $R$ filter in 2002 and 2003
with the 1.5 m Russian-Turkish joint telescope (RTT150) at the TUBITAK
(The Scientific and Technical Research Council of Turkey) National Observatory in Antalya Turkey,
as part of a large program on the 
CCD photometry of Cataclysmic Variables (CVs).
We report the detection of two distinct periodicities in the
light curve of the nova : a) P$_1$=0.31449(15) d -- 7.6 h,   
b) P$_2$=0.017079(17) d -- 24.6 min. 
The first period is evident in both 2002 and 2003 whereas the
second period is only detected in the 2003 data set. 
We interpret the first period as the orbital period of the system
and attribute the orbital variations to aspect changes of the secondary
irradiated by the hot WD. We suggest that 
the nova was a Super Soft X-ray source in 2002 and, perhaps, in 2003. 
The second period could be a QPO originating from the 
oscillation of the ionization front (due to a hot WD) below the inner Lagrange point as predicted by King (1989)
or a beat frequency in the system as a result of the magnetic nature of the WD if steady accretion 
has already been re-established.

\end{abstract}


\section{INTRODUCTION}

Classical novae are a subset of cataclysmic variables which are interacting
binary systems hosting a main-sequence secondary (sometimes a slightly
evolved star) and a collapsed primary component, a white dwarf (Warner 1995).
An outburst on the surface of the white dwarf as a result of a thermonuclear
runaway in the accreted material causes
the ejection of 10$^{-3}$ to 10$^{-7}$ M$_{\odot}$ of material at velocities
up to several thousand kilometers per second (Shara 1989; Warner 1995).  
V2275 Cyg (Nova Cyg 2001 No.2) was discovered  at a magnitude
7.0-8.8 on 2001 August 18 simultaneously by Nakamura et al. (2001) and
Nakano et al. (2001). Early optical spectroscopy showed hydrogen Balmer lines
with P Cygni profiles and H$\alpha$ lines indicating expansion
velocities of 1700 km s$^{-1}$ (Ayani 2001).  
At later stages, high energy coronal lines were found to dominate the
spectrum (e.g., [Si X], [Si IX] and [Al IX]). The nova was found to
belong to the "He/N" subclass of novae defined by Williams (1992), because of
the broad lines of H, He and N in its spectrum (Kiss  et al.
2002). In addition, Kiss et al. (2002) measured t$_2$=2.9$\pm$0.5 d, t$_3$=7$\pm$1 d
and M$_V$= -9$^{m}$.7$\pm$0.$^{m}$7 which were used to derive a distance of
3-8 kpc for the nova. 
A USNO star of $R$=18$^{m}$.8 and
$B$=19$^{m}$.6 was suggested as a possible progenitor star (Schmeer et al. 2001).

This paper is on the CCD photometry of V2275 Cyg covering years 2002 and 2003
obtained with the 1.5 m Russian-Turkish joint telescope (RTT150) at the TUBITAK
National Observatory (TUG) in Antalya, Turkey.  
We present the discovery of a distinct period reported by
Balman et al. (2003) (which we mark as P$_1$) and the detection of a highly coherent QPO 
(Quasi-periodic oscillations) on
6 different nights in 2003 (using $V$, $R$, $I$ filters). Fast variations from V2275 Cyg 
similar to  this second period (which we mark as P$_2$) were also reported by Garnavich
et al. (2004) derived from a data set of two nights obtained with a $V$ filter.

\begin{figure*}
\begin{center}
\epsfig{file=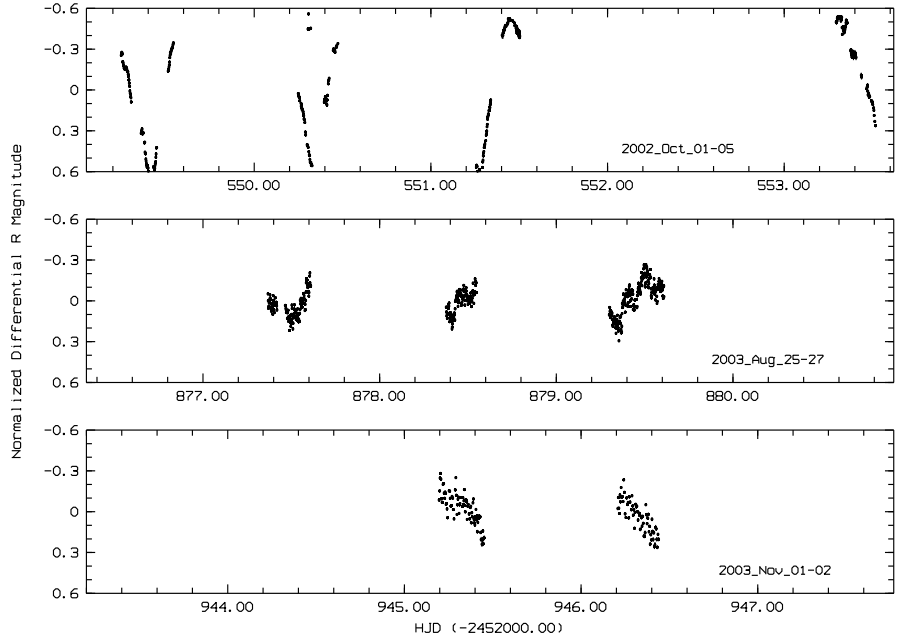,width=9.0cm,height=12.7cm,angle=-90}
\end{center}
\caption{The top panel presents the normalized differential light curve of V2275 Cyg
observed on 2002 October 01-05. The middle panel displays the
light curve obtained on 2003 August 25-27 and the bottom
panel shows the light curve obtained on
2003 November 01-02.
All data are taken with the TUG 1.5 m telescope
using the standard R band filters (Johnson and Cousins). The epoch of the
observations are noted on the x axis.
The average magnitude errors are $\pm$0.0083, $\pm$0.0065 and $\pm$0.0095
for the top, middle and bottom panels, respectively.}
\end{figure*}    

\begin{table}
\center{
\caption{The Time Table of the Observations -- (All observations
are obtained with standard $V$, $R$, $I$ filters)}
\begin{tabular}{llllll}
\hline
UT & Time of Start  & Run &  Number of  & Filter\\
Date & (HJD-2452000) & Time (hr) & Frames  &
\\
\hline
020610 & 436.484518  & 1.8 & 30 & R 
\\
020612 & 438.486060  & 1.9 & 29 & R
\\
021001 & 549.246533  & 7.1 & 88 & R
\\
021002 & 550.248770  & 5.3 & 68 & R
\\
021003 & 551.256457  & 5.9 & 99 & R
\\
021005 & 553.293507  & 5.4 & 80 & R
\\
021201 & 610.238175  & 1.5 & 40 & R
\\
021222 & 631.188612  & 2.9 & 68 & R
\\
030825 & 877.367836  & 5.8 & 136 & R
\\
030826 & 878.376911  & 4.1 & 114 & R
\\
030827 & 879.299621  & 7.5 & 204 & R
\\
031101 & 945.196586  & 6.2 & 81,82 & R,I
\\
031102 & 946.208050  & 5.6 & 65,64 & R,V
\\
031228 & 1002.221443  & 2.0 & 23,23 & I,V
\\
\hline
\end{tabular}
}
\end{table}   

\section{OBSERVATIONS AND DATA REDUCTION}

V2275 Cyg is observed during 8 nights in 2002 and 6 nights in 2003 using
the standard $R$ filters (Johnson and Cousins)
at TUG (see Table 1 for a time table of observations).
The data are obtained with the
imaging CCD (a Loral LICK3 2048$\times$2048 pixels back illuminated CCD chip
at 0.26 arcsec/pix resolution) on 2002 June 10, 2002 June 12 and
Ap47p CCD (1024$\times$1024 pixels
with 13 microns/pix resolution) on 2002 October 1-5, 2002 December 01, and 2002 December 22.
The rest of the data are taken
with the ANDOR CCD (2048$\times$2048 pixels at 0.24 arcsec/pix resolution) on
2003 August 25-27
and the imaging CCD on 2003 November 1-2.
The exposure time is 90 seconds for each frame.
A total of 1149 images are obtained in the R band and reduced using standard procedures
calibrating the frames with the bias/dark current frames and dome flat fields.
In addition, a total of 105 frames in the $I$ band and  87 frames in the $V$ band
have been compiled for comparison (in 2003 November 1-2; in 2003 December 28). 
After the raw data are cleaned and calibrated,
the instrumental magnitudes of the nova are derived by the PSF fitting algorithms DAOPHOT 
(Stetson 1987) and ALLSTAR in the MIDAS software package 
(version 02FEBpl1.0 and 03FEBpl1.0) using 25
stars as PSF stars. Also, another reference group of four constant stars close
to the nova in the same field is formed in order to reduce the scintillation effects and to derive
the relative magnitudes. 
The calibrated apparent magnitude of the nova in the R band
varied in the range from 15$^{m}$.1 to 16$^{m}$.2 in 2002. The fading of the nova continued, 
changing this range to 17$^{m}$.3--19$^{m}$.0 in 2003.

\section[]{ANALYSIS AND RESULTS}

Using our reduced and calibrated $R$ band data,
we constructed light curves for the given nights in Table 1.
A collection of normalized light curves obtained from
runs that are longer than 5 hrs is displayed in Fig. 1.
The top panel is of 2002 October 01-05, the middle
panel is of 2003 August 25-27 and the bottom panel is of
2003 November 01-02. A Deep modulation of the light curve can  be
seen in Fig. 1. Figure 2 shows 
the data obtained in the longest run (7.5 hr) on 2003 August 27. This figure, not only
reveals  the long period that is causing the deep modulation (which we mark as P$_1$), but also  the other sumperimposed
faster variations (which we mark as P$_2$) and humps that are observed in the system.
\begin{figure}
\begin{center}
\epsfig{file=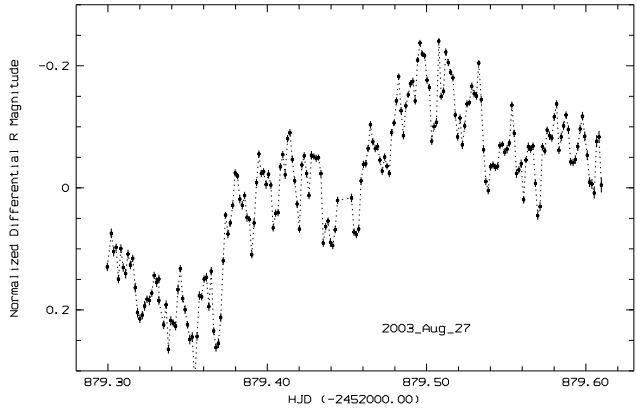,width=5.9cm,height=9cm,angle=-90}
\end{center}
\caption{The light curve of the 
longest observing run (7.5 hr) on 2003 August 27.
The data are taken with the ANDOR CCD
using a standard R band filter (Cousins). The average magnitude error is $\pm$0.0063.}
\end{figure} 
\begin{figure}
\begin{center}
\epsfig{file=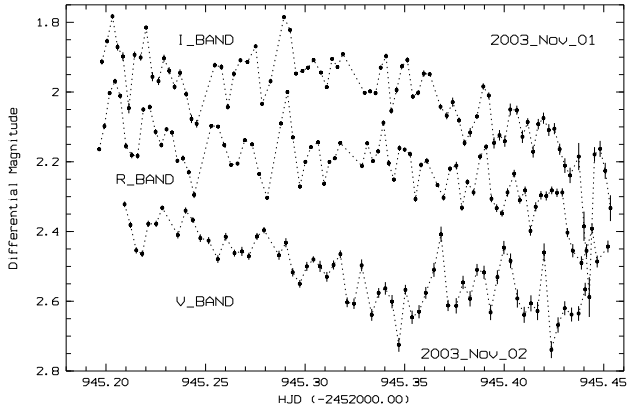,width=5.9cm,height=9cm,angle=-90} 
\end{center}
\caption{The light curve of V2275 Cyg obtained with the standard $I$, $R$ filters
on 2003 November 01, and $V$ filter on 2003 November 02. 
The data are taken with the imaging CCD.
The average magnitude errors are $\pm$0.015, $\pm$0.008,
and $\pm$0.011 for the $V$, $R$, $I$ light curves, respectively.}
\end{figure}
We also accumulated  data using the standard
$I$ filter on 2003 November 01 and $V$ filter on 2003 November 02 together with the $R$ filter
observations (see Fig. 3).
The short timescale variations (which we mark as P$_2$)
are apparent in the figure.
Subtraction of a linear trend from the $V$ and $I$ band light curves in Fig. 3
suggests that the long period variations (i.e., P$_1$) are 
larger in amplitude in the $I$ and $R$ bands than the $V$ band which can be 
expected (see Discussion).
\begin{figure}
\begin{center}
\epsfig{file=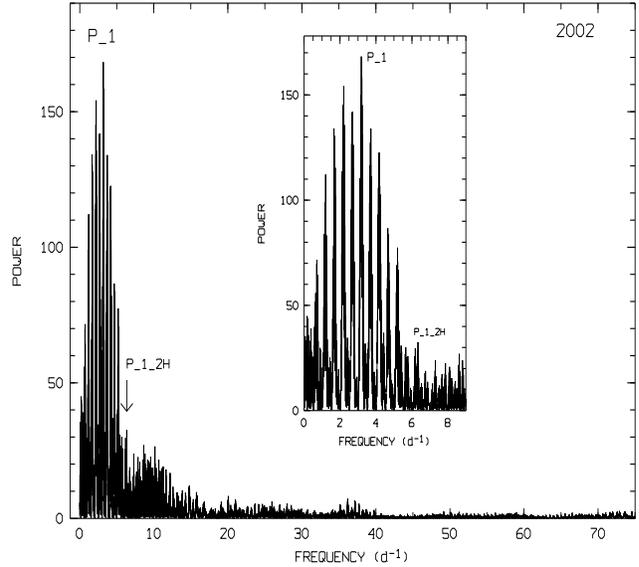,width=8cm,height=9cm,angle=-90}
\end{center}
\caption{The power spectrum of
V2275 Cyg, derived from the 2002 data set (8 nights), using the Scargle Algorithim.
The new period is indicated as P$_1$ and its weak (but significant) second harmonic
is also noted.
}
\end{figure}
\begin{figure}
\begin{center}
\epsfig{file=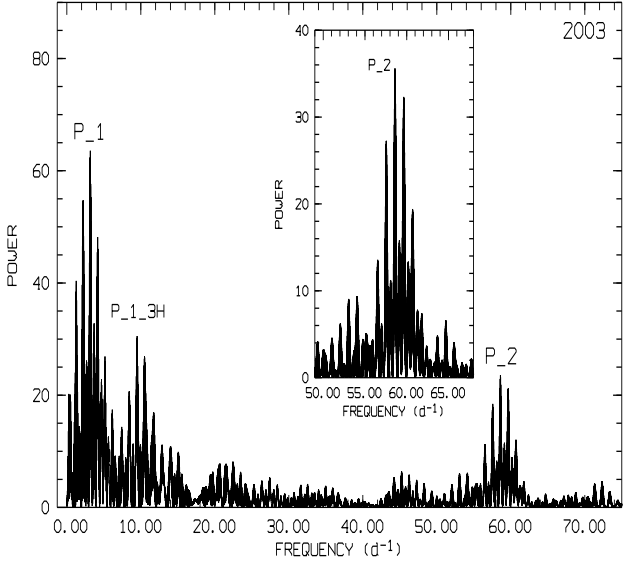,width=10cm,height=10cm,angle=-90}
\end{center}
\caption{The power spectrum
of V2275 Cyg obtained using the 2003 data set (5 nights).
P$_1$ is noted on the figure.
The inset panel is a normalized
power spectrum of V2275 Cyg in the frequency range centered around
P$_2$ where P$_1$ is removed.
A Scargle algorithim is used for the analyses.
}
\end{figure}   

We performed fourier analysis of the
time series obtained from the data in order to derive the periods of these 
modulations. In general, several standard programs have been used
like the Scargle Algorithim (Scargle 1982) and Discrete Fourier
Analysis using Leahy normalization (Leahy et al. 1983). Figure 4 shows the power spectrum of the data
for the year 2002 and Fig. 5 for the year 2003 where the Scargle algorithm is used to
calculate the power spectra.
The detection limit of a period at the 3$\sigma$ confidence level (99$\%$) is
a power of 14.2 in Fig. 4 (2002 data) and 15.6 in Fig. 5 (2003 data) (see also Scargle 1982).
In order to correct for the effects of windowing and sampling functions on power spectra,
synthetic constant light curves are created and a  few very prominent frequency peaks that appear
in these light curves are prewhitened from the data in the analysis. Before
calculating the power spectra, the individual or consecuitive nights are normalized by
subtraction of the mean magnitude. Moreover, when necessary, the red noise in the lower frequencies
is removed by detrending the data.

We found a prominent period at P$_1$=0.31449(15) d using the whole data set.
The power spectra in Fig. 4 and Fig. 5 
show the highest peak at this period
and the group of peaks around
it are some of the $\pm$1/3, $\pm$1/2, $\pm$1, and $\pm$2 day 
aliases of the detected period. 
In Fig. 4 (2002 data) a weak second harmonic of P$_1$ (P$_1$/2) is present (it is significant).
In Fig. 5 (2003 data) the third harmonic 
of P$_1$ (P$_1$/3) is also present, but the second harmonic is not significantly detected. 
The period (P$_1$) shows an amplitude variation of
0.${^m}$42$\pm$0.${^m}$06 in 2002 (measured by fitting a sine wave).  
The amplitude of the variations are decreased significantly in 
2003 to 0.${^m}$22$\pm$0.${^m}$12 .
The decrease in modulation depth is about 50$\%$ (in magnitude).
The ephemeris for P$_1$ determined by fitting a sine curve are :\\

\noindent
T$_{min}$= HJD 2452549.4163($\pm$0.0154) + 0.31449($\pm$0.00015)E

\vspace{0.3cm}

We also detect a second periodicity at P$_2$=0.017079(17) with 
an amplitude of 0.${^m}$03$\pm$0.${^m}$01. 
These rapid variations are revealed
in all the nights in 2003 with varying intensity. 
They are also clearly seen in Figs. 2 and 3 (in the $V$, $R$, $I$ bands). 
We do not recover 
the beat period between P$_2$ and P$_1$ in the 2003 data set. 

Figures 6 and 7 display the mean light curves folded on P$_1$ using the 2002 
and 2003 data set, and
P$_2$ using the 2003 data set, respectively. 
The shape of the profiles are sinusoidal for both
P$_1$ and P$_2$ (in 2002-2003) with a slight asymmetry. Hump-like features
are apparent on both profiles, as well. 
The ingress in the mean light curve 
of P$_1$ and P$_2$ is shorter by 10$\%$ of the photometric phase than the egress.
In adition, we have constructed color diagrams $V-R$, $I-R$, and $I-V$ to search for color variations
over the photometric phases of the periods (light curves are calibrated before the construction
of the color diagrams).  
The right-hand panel of Fig. 8 shows the existence of 
a significant modulation of the
color difference $I-V$ over the photometric phase of P$_1$. 
The $I-R$ magnitudes plotted on the 
left-hand panel of Fig. 8 indicates color variation over the photometric phase of the
second period as well. 
\begin{figure}
\begin{center}
\epsfig{file=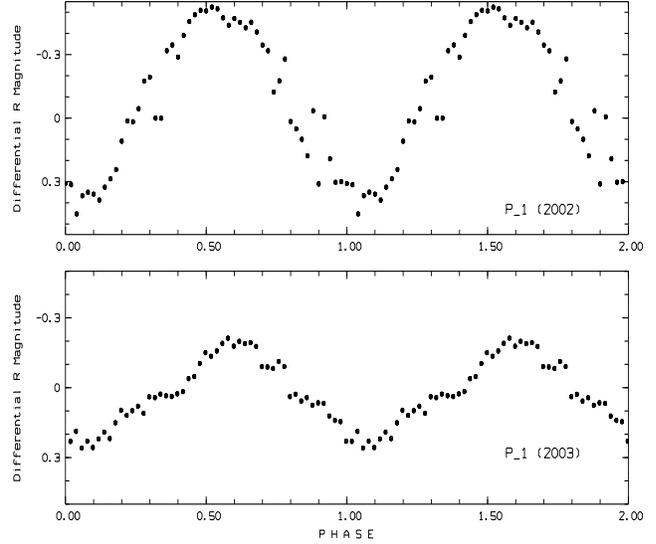,width=7.8cm,height=8.8cm,angle=-90}
\end{center}
\caption{The light curve of the 2002 data set, top panel, and 2003 data set, 
bottom panel, folded on the period P$_1$ [0.31449(15)]. 
The first data point in time (start HJD in 2002) 
is taken as the reference and a grouping (averaged over) of 50 phase bins is used for the folding process 
(i.e., the 2002 and 2003 data are phase-locked). 
The average error of a phase bin is $\pm$0$^m$.0021 for the top and $\pm$0$^m$.0024 for the
bottom panel.}
\end{figure}
\begin{figure}
\begin{center}
\epsfig{file=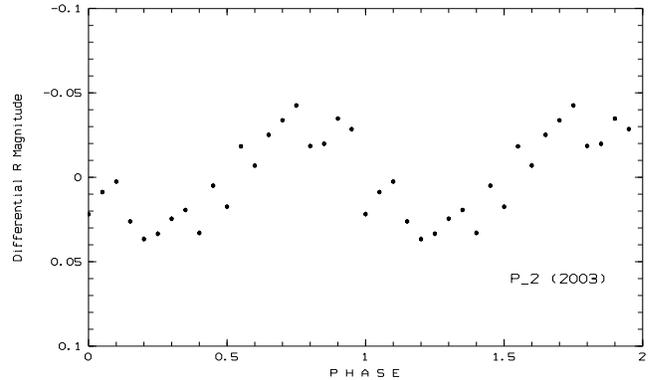,width=5.6cm,height=9cm,angle=-90}
\end{center}
\caption{The light curve of the 2003 data set, folded on the
period P$_2$ [0.017079(17)]. The first data point in time (start mid-HJD in 2003)
is taken as the reference and a grouping (averaged over) of 20 phase bins is used for the folding 
process. The average error of a phase bin is $\pm$0$^m$.0015 . The small 
oscillations superposed on the
mean light curve are due to the time windows of the data.}
\end{figure}
\begin{figure}
\begin{center}
\epsfig{file=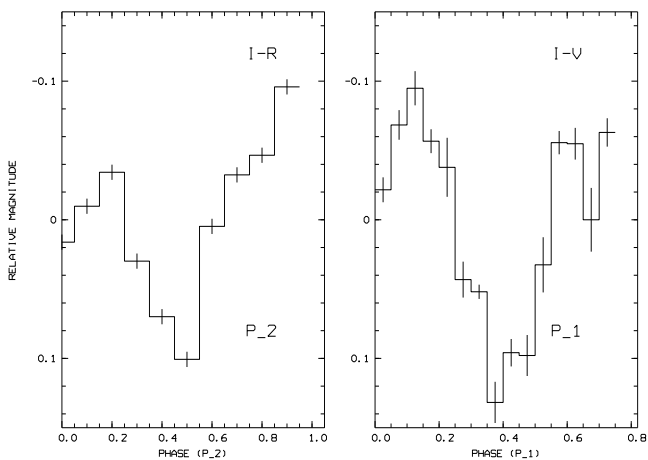,width=6.7cm,height=9cm,angle=-90}
\end{center}
\caption{Color variations of the detected periodicities. The left-hand panel
is the
normalized color magnitude ($I-R$) curve vs. photometric phase of P$_2$.
The total lightcurve is
folded on  P$_2$ using 10 phase bins. 
The right-hand panel shows $I-V$ color variation
over the photometric phase of P$_1$. The total lightcurve is
folded on  P$_1$ using 15 phase bins.} 
\end{figure}
\section[]{DISCUSSION}

We presented 14 nights of data on V2275 Cyg obtained with the TUG 1.5 m telescope using
mainly standard $R$ filters in 2002 and 2003. 
We discover large modulations $\Delta m_r$=0.42 at P$_1$=0.31449(15) d 
in the light curve of the classical nova in 2002 and the amplitude of this modulation
decrease to $\Delta m_r$=0.22 in 2003. This period 
is discovered in the $V$ and the $I$ band as well. Since the periodicity is persistent and coherent,
we propose that this is the
binary period of the system. The orbital period can be  detected as a result of
the aspect variations of the secondary due to heating from the hot WD 
(Kovetz, Prialnik, $\&$ Shara 1988). The decrease in the modulation depth
from 2002 to 2003 and the color variations ($I-V$, $I-R$, $R-V$)  support a scenario where 
the heated secondary is the source of the light and the color variation rather than 
a hot spot on the disk.     
The Mass-Period relation M$_2$$\simeq$0.11P$_{hr}$ (Warner 1995) yields a secondary mass of
about 0.83 M$_{\odot}$ for V2275 Cyg using the discovered binary period in this paper, 
which is in accordance with the fact that a massive secondary will be hot and prone to 
irradiation effects.
In addition, the mean light curves (Fig. 6 and 7)
indicate an asymmetry in the R band as would be expected 
from the differential rotation of the secondary spreading the
heated atmosphere into a nonspherical shape (tear-drop model). 
An eclipse can be ruled out because the modulation amplitude would be larger in time as observed
for some other novae (e.g., V838 Her: Leibowitz et al. 1992, V1494 Aql: Kato et al. 2004). 
A recent search for variations/periodicities in the light curve of faint CVs (including old novae)
reveals that irradiation yields larger amplitude for modulations (Woudt $\&$ Warner 2003a,b)
as detected in our study.

After a nova explosion, the hot WDs are candidates for heating their cooler companion.
Some classical nova systems were recovered to show this irradiation effect like: 
(1) V1500 Cyg (N Cyg 1975)
which showed an unperturbed temperature of 3000 K for the secondary and 8000 K for the heated side 
(Schmidt et al. 1995; Somers $\&$ Naylor 1999); (2) DN Gem (N Gem 1912)
(Retter, Leibowitz $\&$ Naylor 1999); (3) WY Sge (N Sge 1783) which indicated that the 
accretion luminosity from the disk could even be responsible for the irradiation of the secondary 
(Somers, Ringwald $\&$ Naylor 1997).
However, only one of such systems resemble V2275 Cyg closely:
orbital  modulations due to aspect variations of the heated face of the 
secondary was detected for Nova Cygni 1992 (V1974 Cyg) at the outburst stage (about a year 
and a half after outburst) 
by DeYoung $\&$ Schmidt (1994) before the WD cooled to a point where it was no longer a strong X-ray source. 
The orbital variations in V1974 Cyg were $\sim$0.1 in the $I$ band and $<$0.05 in the $V$ band
when the H-burning had just turned-off (the WD temperature was about
3-4$\times$ 10$^5$ K; Balman et al. 1998).
V1974 Cyg was discovered as a Super Soft X-ray source (SSS) while burning the hydrogen
over its surface (Balman et al. 1998 and references therein) consistent
with the fact that
most novae are expected to be a SSS at a certain point during their outburst stage
(see Krautter 2002 for a review). 
A hot WD at a temperature above 1$\times$10$^5$ K (i.e., emits mostly in the soft X-ray wavelengths)
with a radiative wind can be the source of an ionization front 
irradiating the secondary star. 
Therefore, we suggest that V2275 Cyg was also a SSS during 2002-2003, and the
radiative winds/ionization front of a SSS reaching out well into the 
photosphere of the secondary, can explain the observations of  V2275 Cyg in 2002. 
The decrease of 50$\%$ in the modulation
depth from 2002 to 2003 is another indication of the changing conditions in the ionization front.
The strong third harmonic of P$_1$ detected in 2003 could support the existence of  several hot zones over the
total surface of the secondary along with the significantly heated inner face resulting in the higher harmonics
becoming prominent.

We also detected another periodicity P$_2$=0.017079(17) d in the light curve of 2003.
We suggest that this is a  highly coherent QPO from the system. The power of the signal
varies on different nights (independent of length of the observation 
together with the small changes of the periodicity which differ according to
the color (i.e., filter). The characteristics of the oscillations revealed in 2003 
does not strongly support a WD-spin period as the origin. 
The timescale and characteristics  of the QPOs ($\sim$1475 s)
resemble flickering, or reprocessing from blobs orbiting within the inner regions/magnetosphere
of the accretion disk in accreting CVs. It could also be the beat period between the spin period
and the orbital period of the system (P$_1$).
However,   
since the nova is still in its early outburst stage during the TUG observations, 
it is not clear whether the disk is completely
disrupted, or if  re-established, the accretion is steady, sporadic or unstable. 
For example, the existence of an accretion disk was revealed in V1974 Cyg two years
after the outburst (Retter et. al 1997) which was long after 
the discovery of the variations due to irradiation of the secondary.         

There have been recent detections of oscillations in classical nova systems
of about 2500 s (V1494 Aql: Drake et al. 2003)
and $\sim$ 1300 s (V4743 Sgr: Ness et al. 2003) in the X-ray wavelengths. 
These oscillations are attributed to expected WD pulsations 
(i.e., nonradial nonadiabatic modes) from the
$\sim$1.5$\times$ 10$^5$ K and L$\sim$2000L$_{\odot}$, hot WDs (Starrfield et al. 1985).
The P$_2$ detected in V2275 Cyg 
is similar to these in timescale. 
However, it is expected that the higher luminosity (a factor of 10) and the 
temperature (a factor of 4-5) of the WDs, as the H is burned over the surface, 
should decrease the pulsational timescale (Starrfield et al. 1985), which makes this
scenario unfavorable for V2275 Cyg.
In addition, the 2002 data set should have revealed the same oscillations, if they were solely due to
WD pulsations. 

A more plausible explanation for the origin of P$_2$ is, yet again, the effect of an ionization front
 (IF) resulting from an alleged SSS phase and the radiative winds of the WD in the outburst stage. 
If the IF reaches well into the photosphere of the secondary, favorable conditions
could result in ionization over the face of the secondary (IF penetrates photosphere) resulting in 
"boiling off" (may be "ripping off") material from a dense static/neutral medium which could be the case
in 2002. If IF does not reach as far as the photosphere, it could still reach out below the
inner  Lagrangian point (L$_1$ : a sonic point) where conditions can not be steady and the IF then oscillates
on a timescale of the order of the dynamical timescale of the secondary near L$_1$ which could
result in QPOs as suggested by King (1989). This oscillation time scale is 
proportional to t$_{qpo}$$\simeq$H/c$_s$ (scaleheight/sound-speed) which is
equivalent to t$_{qpo}$$\simeq$(R$^3$/GM)$^{1/2}$$\simeq$176 P$_{hr}$. Given the orbital period
derived in this paper (P$_1$), the predicted oscillation period of L$_1$ is similar to P$_2$ (assuming
the size R of the secondary will be larger on the equatorial plane). 
By late 2003 the IF could be withdrawn toward the WD irradiating only the
inner Lagrangian point (L$_1$) which yields the QPOs. 

\section[]{CONCLUSIONS}

We detected two periodicities in the light curve of the classical nova V2275 Cygni (2001).
The first one is 0.31449(15) d which we attribute to the orbital period of the binary system.
The modulation depth of the mean light curve folded on this period indicates a change from 0$^m$.42$\pm$0$^m$.06  to 
0.${^m}$22$\pm$0.${^m}$12 over the course of one year. 
We propose that these variations are due to the illumination of the
secondary by the hot WD where the aspect variations of the heated/irradiated secondary
reveals the binary period of the system. The reduction in the modulation depths indicates the 
changing conditions in the ionization front at the location of the secondary. This highly ionizing
radiation could originate from a hot WD which went through a Super Soft X-ray phase while burning
H on its surface. We also detect a second period  0.017079(17) d  in the year 2003 which we interpret as a QPO from 
our data in 2002-2003. For the origin of this period 
we propose either a magnetic cataclysmic variable scenario  
(e.g, the interaction of blobs with the magnetosphere) or a scenario where the QPOs are a result of
oscillations of the inner lagrangian point (L$_1$) due to irradiation from a cooling hot WD.
Both scenarios indicate that some form of accretion is being established in the 
nova system after the outburst.
Our preliminary analysis reveal that both periods exist in 2004 data and the modulation depth of
the first period is further reduced.
The photometric observations of V2275 Cyg in 2004 (still conducted at TUG) should reveal the changing shape
and depth of the orbital modulations and QPOs which  will, inturn, portray the evolution of the WD itself.

\section*{Acknowledgments}

We would like to thank an anonymous referee for critical reading of the
manuscript]. We also thank the TUBITAK National Observatory Institute and 
\"U. K{\i}z{\i}lo\u{g}lu for the support of the observations. 
We thank Tansel Ak for obtaining data for us. SB
also acknowledges the support by the Turkish Academy of Sciencies TUBA-GEBIP 
fellowship (Distinguished Young Scientist Award). AR was partly supported by 
a postdoctoral fellowship from Penn State University.

\label{lastpage}

\end{document}